\documentclass{aa}
\usepackage{txfonts}
\usepackage{graphicx}
\usepackage[]{natbib}

\begin{document}

\title{Large-scale environments of $z<0.4$ active galaxies}
\author{H. Lietzen \inst{1} 
 \and P. Hein\"am\"aki \inst{1} 
 \and P. Nurmi \inst{1}
 \and L. J. Liivam\"agi \inst{2 \and 3}
 \and E. Saar \inst{2}
 \and E. Tago \inst{2}
 \and L. O. Takalo \inst{1}
 \and M. Einasto \inst{2}}

\institute{Tuorla Observatory, Department of Physics and Astronomy, 
 University of Turku, V\"ais\"al\"antie 20, 21500 Piikki\"o, Finland
 \and Tartu Observatory, 61602 T\~oravere, Tartumaa, Estonia
 \and Institute of Physics, University of Tartu, T\"ahe 4, Tartu 51010, Estonia}

\date{Received / Accepted }

\abstract
{Properties of galaxies depend on their large-scale environment. As the influence of active galactic nuclei (AGN) in galaxy evolution is becoming more evident, their large-scale environments may help us understand the evolutionary processes leading to activity. The effect of activity can be seen particularly by showing whether different types of active galaxies are formed by similar mechanisms.}
{Our aim is to study the supercluster-scale environments of active galaxies 
up to redshift 0.4. Our data include quasars, BL Lac objects, Seyfert, and radio galaxies.}
{We used a three-dimensional, low-resolution luminosity-density field constructed of a sample of luminous red 
galaxies in the seventh data release of the Sloan Digital Sky Survey. We calculated the average density of this field in a volume of a 3\,$h^{-1}$Mpc sphere around each AGN for estimating the environmental density levels of different types of AGN. This analysis gives us the distribution of AGN in the global environment of superclusters, filaments, and voids.}
{Our results show that, while radio-quiet quasars and Seyfert galaxies are mostly located in low-density regions, radio galaxies have higher environmental densities. BL Lac objects usually have low-density environments, but some of them are also in very high-density regions.}
{Our results give support to the theory of galaxy evolution where galaxies are affected by two modes of AGN feedback: quasar mode that turns a blue star-forming galaxy into a red and dead one, and radio mode that regulates the growth of massive elliptical galaxies. We conclude that quasars are in lower density environments than radio galaxies, most likely because the galaxies in rich environments have evolved to a state suitable for radio-loud activity earlier. Galaxies in poor environment have evolved more slowly, and are still going through the earlier quasar-mode feedback in galaxy evolution.}


\maketitle

\section{Introduction}

Groups and clusters of galaxies are not distributed in space randomly, but form a network of superclusters, filaments, and voids \citep{Joeveer1978, Tarenghi1978, Gregory1978, Zeldovich1982}. In the cosmic density field, voids represent large underdense regions, while superclusters constitute the highest density enhancements in the universe. From the dynamical point of view, the evolution in the void regions is slow and ends early, while in superclusters the dynamical evolution starts early and continues until the present day \citep{Einasto2005}.

Superclusters consist of a wide variety of galaxy systems ranging from individual galaxies to rich clusters of galaxies aligned to filaments. Superclusters are found to contain diverse environments of virialized systems and diffuse matter with differing ratios of dark and optically visible matter \citep{Proust2006, Gilmour2007}. In addition, numerical simulations and theoretical calculations by \citet{Cen1999} predict that superclusters are large-scale reservoirs of so-called missing baryons. The XMM-Newton follow-up program for validating Planck cluster candidates report the detection of a supercluster through the Sunyaev-Zeldovich effect \citep{Sunyaev1972}, indicating the distribution of gas on very large scales \citep{Planck2011}. 

On cluster scales, it is well known that the density of the environment affects the properties of galaxies. The effect is detected as the morphology-density relation \citep{Dressler1980} and in the star-formation rates \citep{Gomez2003}. Luminous galaxies are also in general more clustered than the faint ones \citep{Hamilton1988}. 

It is increasingly evident that the global environment (supercluster-void network) also plays an important role in determining galaxy properties. Also, galaxy activity and star-formation rate seem to be influenced by the large-scale environment in which they reside \citep[e.g.][]{Balogh2004, Einasto2005, Gao2005, Gilmour2007, Porter2008, Skibba2009, Lietzen2009, Wang2011}.  For example, galaxy morphology in superclusters seems to depend on the richness of the supercluster. According to \citet{Einasto2007b}, there are more early-type galaxies in rich superclusters than in the poor ones. Galaxies in the cores of superclusters are also more likely to be red, early-type galaxies, while blue late-type galaxies dominate the ourskirts of the superclusters. \citet{Einasto2008} found that in equally rich groups there are more early-type galaxies in supercluster cores than in the outskirts of superclusters. Also, the galaxies that do not belong to any group are more often early-type if they are in the supercluster cores. \citet{Tempel2009,Tempel2011} show that the luminosity function of elliptical galaxies depends strongly on global environment, indicating that the global environmental density plays an important role in the formation history of elliptical galaxies. Spiral galaxies are not as sensitive to environmental density as elliptical galaxies, suggesting that the formation mechanism of spiral galaxies is almost independent of the large-scale environment. 

Presumably the properties and evolution of galaxies are closely connected to an interplay between galaxies and their environments on different scales. While the period of active galactic nuclei (AGN) may be a short but important phase in the evolution of all massive galaxies \citep{Cattaneo2009}, an interesting question is how the different types of AGN activity relate to the global environment. Based on this background, the goal of this paper is to study the environments of AGN on supercluster scales. 

The standard model of an AGN assumes that the energy is produced by the accretion of matter into a supermassive black hole surrounded by a dusty torus. The unification scheme connects the different types of AGN in this model, stating that the differences in observational properties are caused by viewing the targets from different directions \citep{Antonucci1993}. According to the unification model of radio-quiet AGN, Seyfert 2 galaxies are observed through the obscuring dust torus, while radio-quiet quasars and Seyfert 1 galaxies are seen without obscuration. For the case of radio-loud AGN, \citet{Urry1995} outline two separate unification schemes: BL Lac objects are viewed from the direction of the jets of low-luminosity radio galaxies, while radio-loud quasars are more luminous radio galaxies viewed from the direction of their jets. Radio luminosity divides radio galaxies in two morphologically different classes. High-luminosity radio galaxies belong to the FR II type, while the low-luminosity radio galaxies have FR I type morphology. In FR I type galaxies the intensity gradually declines when moving farther from the host galaxy. FR II galaxies, on the other hand, are edge-brightened and have luminous hotspots at the outer edges of their radio emission \citep{Fanaroff1974}.

While orientation of the viewing angle can explain the observed differences of some types of AGN, it is obvious that there are also physical differences in some cases. Different types of AGN activity may represent different phases of galaxy evolution. The evolutionary model presented by \citet{Hopkins2008} states that galaxy mergers cause gas inflows, which then lead to the growth of the supermassive black holes. These mergers trigger the activity observed as quasars. Activity may also be triggered by secular processes, e.g. bar or disk instabilities. These processes are likely to produce lower luminosity Seyfert galaxies \citep{Hopkins2007}, but quasars can also form without a major merger \citep{Bournaud2011}. 

\citet{Croton2006} add another type of AGN feedback to the models of galaxy evolution: besides the merger-driven `quasar mode', galaxies are also affected by lower energy `radio mode'. The radio mode starts working when gas from a static hot halo flows to the center of a galaxy feeding its supermassive black hole. The radio mode affects the galaxy by heating the gas, thereby preventing cooling. While the quasar mode is responsible for stopping the star formation and turning the galaxy into a red and dead one in a short time, the radio mode warms the surroundings of the elliptical galaxy and regulates its growth with a longer timescale. \citet{Croton2006} do not make any prediction on the nature of radio feedback. However, \citet{Fanidakis2011} show that jets can cause suitable feedback and explain the observed luminosities of AGN. 

There is very little connection between the theories based on the semi-analytic models that we described above and the observed AGN. For example, \citet{Cisternas2011} find no connection between AGN activity and merger remnants: the rate of morphological disturbances in AGN hosts was low and similar to that of normal galaxies. Similar results have also been found by \citet{Gabor2009} and \citet{Georgakakis2009}, both of whom conclude that AGN are not likely to be triggered by mergers.  Observing signs of a recent merger in an AGN host galaxy is difficult, since the active nucleus is very bright compared to the galaxy. Studying the large-scale environments is an indirect way of finding a connection between evolutionary models and the observed AGN population. It is also possible that the environment affects the triggering of AGN through different gas content of the surrounding intergalactic medium instead of mergers between galaxies.

Environments of different types of AGN have been studied earlier on different scales \citep[e.g.][]{Padmanabhan2009,Donoso2010,Hickox2009,Bornancini2010}. Most of the previous studies have concentrated on the environments on a group or cluster scale. In one of the first studies dedicated to the large-scale structure of the Universe, \citet{Joeveer1978} found that radio galaxies are abundant in superclusters. In a more recent work, \citet{Gilmour2007} have found the optically selected AGN are not found in the densest regions of supercluster cores. The same is found also by \citet{Kocevski2009} in supercluster CL1604. In this work, we study the environments of different types of AGN on a large scale.

In \citet{Lietzen2009} the environments of quasars were studied using the SDSS DR-5 galaxy and group catalogs. The main result was that the nearby quasars avoid the densest areas, and are more likely to be found at the edges of the superclusters. The DR-5 galaxy data reached the distances of 500\,$h^{-1}$Mpc, and there were 174 quasars in the same volume. To take full advantage of the current observations we extend our analysis in this paper to the SDSS seventh data release (DR-7). To increase the volume of our sample we concentrate on the sample of luminous red galaxies (LRGs), which is an extension of the main sample of galaxies. 

As red elliptical galaxies, LRGs are typically the most luminous galaxies in the Universe.  According to the morphology-density relation, they are usually concentrated in the centers of clusters \citep{Dressler1997}. This makes them good indicators of clusters of galaxies and the large-scale structures \citep{Gladders2000}. Because LRGs are more luminous than normal galaxies, they can be detected from greater distances. This makes it possible to study the large-scale structure of the universe in a larger volume when using LRGs instead of normal galaxies.

The volume of the LRG sample contains over three thousand quasars, greatly improving the statistics. We constructed a luminosity-density field based on LRGs and calculated the average environmental density around each AGN. In addition to quasars, we now also extend our studies to Seyfert galaxies, BL Lac objects, and radio galaxies. This makes it possible to see if there are differences between the different types of AGN and to test the unification and evolutionary schemes. In this paper we study the distribution of AGN in the global environment of superclusters, filaments, and voids.

In this paper we assume a $\Lambda$CDM cosmology with $\Omega_M=0.27$, 
$\Omega_{\Lambda}=0.73$, and $H_0=100 h$\,km\,s$^{-1}$\,Mpc$^{-1}$.

\section{Data}
\subsection{Density field data}
As a basis we used a luminosity density field of luminous red galaxies to study AGN in their global environments. We used data from the seventh data release (DR-7) of the SDSS \citep{Abazajian2009}. The LRGs were selected by an SQL query requiring the PrimTarget to be either TARGET\_GALAXY\_RED or TARGET\_GALAXY\_RED\_II. Reliable redshifts were required by the conditions SpecClass~$=2$ and zConf~$>0.95$. We used $K$ and evolution corrections by \citet{Eisenstein2001} to calculate the absolute magnitudes. For limiting the errors in magnitudes of the galaxies in our sample, we required their absolute magnitudes to be $M_g>-23.4$.

We also used the SDSS DR-7 main sample of galaxies at $z<0.2$ as a comparison to the LRG sample. Our data contains 583\,362 galaxies in the main area of SDSS DR-7. The magnitudes of the galaxies range from $r=12.5$ to $r=17.77$. The main sample of galaxies is complete to 565\,$h^{-1}$Mpc distance. To suppress the cluster finger-of-god redshift distortions, we used groups of galaxies based on the LRGs that were extracted using the method described in \citet{Tago2010}. The method of constructing the luminosity-density fields based on these data samples is discussed in Sect. \ref{Methods}.

\subsection{AGN data \label{Data}}
\begin{figure}[t]
  \resizebox{\hsize}{!}{\includegraphics{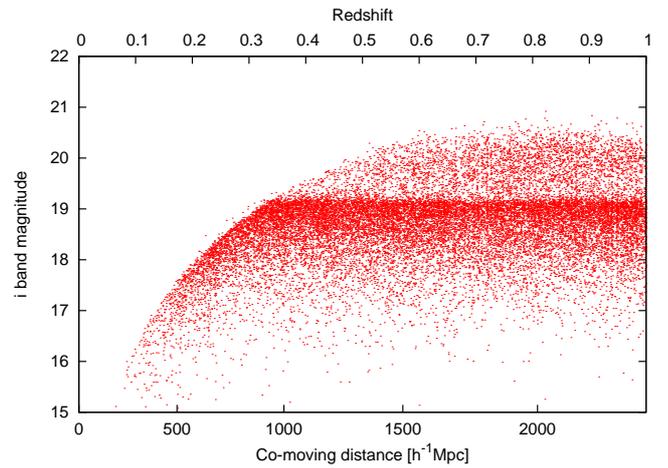}}
  \caption{Relation between the $i$ band magnitudes and redshifts of quasars.}
  \label{zvsmag}
\end{figure}

The quasar catalog by \citet{Schneider2010} contains 105\,783 
quasars. The catalog is based on SDSS DR-7, and it consists of objects with 
luminosities higher than $M_i=-22.0$ that are spectroscopically determined 
quasars. We required our targets to have a spectrum taken as a normal science 
spectrum (SCIENCEPRIMARY\,$=1$) and photometric measurement designated 
PRIMARY in the BEST photometric database. The sample is volume-limited and 
complete to redshift $\sim 0.4$ (corresponding distance 1000\,$h^{-1}$Mpc). 
At greater distances, the limiting magnitude $i=19.1$ starts limiting the 
sample. This can be seen in Fig.~\ref{zvsmag}, where the quasars are plotted by their $i$ band magnitudes vs. redshift. We take this 1000\,$h^{-1}$Mpc for the upper limit in our analysis.
Due to the lack of very nearby quasars, we set a lower distance 
limit for our study at 225\,$h^{-1}$Mpc. To study the differences between radio-quiet and radio-loud quasars, quasars with radio luminosity more than $10^{25}$\,W\,Hz$^{-1}$ in the Faint Images of the Radio Sky at Twenty cm (FIRST) catalog were determined to be radio loud. We use the same limit that was used by \citet{Donoso2010} in order to make our results comparable to theirs.  The flux completeness limit of FIRST is 1\,mJy, which corresponds to a luminosity of $2.2\times10^{23}$\,WHz$^{-1}$ at 1000\,$h^{-1}$Mpc co-moving distance (corresponding luminosity distance 1362\,$h^{-1}$Mpc). This is considerably lower than our luminosity limit for radio-loud quasars. Therefore our sample of radio-loud quasars is complete in terms of radio luminosity. With this constraint, 26 of the quasars are radio loud. Radio-quiet quasars have a radio luminosity less than $10^{25}$\,W\,Hz$^{-1}$ or have no radio detection at all. Our final sample contains 3338 radio-quiet quasars.

For the studies of BL Lac objects, we used the catalog by  
\citet{Plotkin2010}. They present a sample of 723 BL Lac candidates in the 
SDSS DR-7, from which we chose those that are labeled high-confidence BL 
Lac candidates with reliable redshift. We limited 
the sample to the main area of the SDSS and to the distance range of our 
quasar sample. With these limitations, we get a sample of 81 BL Lac objects.

\begin{figure}[t]
  \resizebox{\hsize}{!}{\includegraphics{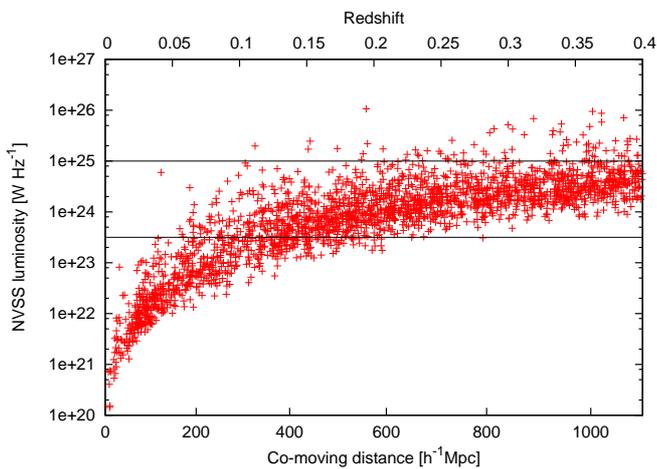}}
  \caption{Radio luminosities of radio galaxies at different redshifts. The lower limit for radio galaxies, $10^{23.5}$\,W\,Hz$^{-1}$ and the limit between FR I and FR II galaxies, $10^{25}$\,W\,Hz$^{-1}$, are marked by lines.}
  \label{radioluminosities}
\end{figure}

Our radio galaxies are from the unified catalog of radio objects by 
\citet{Kimball2008}. We use their sample 5, which contains 
galaxies that have been detected by three radio surveys, 
Faint Images of the Radio Sky at Twenty cm (FIRST), NRAO VLA Sky Survey (NVSS),  Westerbork Northern Sky Survey (WENSS), and the 
SDSS. We calculated the spectral index $\alpha^{92}_{20}$ for galaxies 
in the catalog using the radio luminosities obtained in WENSS and NVSS, and defined samples of steep and flat spectrum radio galaxies 
with a limiting index of $-0.5$, which was also used by \citet{Kimball2008}. Limited between distances 225\,$h^{-1}$Mpc and 
1000\,$h^{-1}$Mpc, we got samples of 1019 steep-spectrum, and 710 flat-spectrum 
radio galaxies. All these radio sources have been classified as galaxies by their optical spectra. The NVSS luminosities of all radio galaxies in the sample are plotted againts redshift in Fig. \ref{radioluminosities}. The flux completeness limit of NVSS is 2.5\,mJy, which corresponds to the luminosity $5.5\times10^{23}$\,WHz$^{-1}$ at the 1000\,$h^{-1}$Mpc comoving distance. For WENSS, the flux limit is 18\,mJy, which corresponds to $3.9\times10^{24}$\,WHz$^{-1}$ at the upper distance limit of our sample. We set a lower limit for NVSS luminosity at $10^{23.5}$\,WHz$^{-1}$ for the radio galaxies used in our analysis. This limit is marked with a line in Fig. \ref{radioluminosities}, which shows that the flux limit makes the sample incomplete at redshifts more than 0.25. We did not set a  limit on FIRST or WENSS luminosities, and the flux limits of these measurements may also cause incompleteness. For these reasons our radio data are not complete and limited in volume. Our method of luminosity-density field handles each target independently of the others, reducing the effects of incompleteness. The only possible problem might be biases caused by luminosity-dependency in the environment. We studied these effects by calculating densities separately for subsamples of radio galaxies with different radio luminosities.

A flat radio spectrum is a sign of spatially concentrated radio emission, which is more typical of quasars than radio galaxies. 
The high number of targets in the radio galaxy sample, which are optically determined galaxies but have flat radio spectra, raises questions as to their nature. Some of these targets could be gigahertz peaked-spectrum (GPS) or compact steep spectrum (CSS) radio sources. These sources typically have powerful compact radio emission that peaks at such low frequencies that their spectra may look flat between 92 and 20\,cm. However, the GPS and CSS are rare objects that cannot explain the majority of flat-spectrum radio galaxies. \citet{Kimball2008} suggest that the flat-spectrum radio galaxies could be radio-emitting Seyfert galaxies or LINERs. This could explain only the faintest end of these galaxies. Some of the flat-spectrum radio galaxies may also be distant FR I type radio galaxies in which the radio emission declines rapidly outside the central parts, resulting in more compact emission with flat spectral index in NVSS and WENSS. In this case we would expect to see some redshift dependency in the fraction of flat-spectrum sources in the radio galaxy sample, which we do not detect.  None of these explanations is expected to produce as high number of flat-spectrum radio galaxies as found by \citet{Kimball2008}. In our study on the large-scale environments, we handled the flat-spectrum radio galaxies as one group, separately from the steep-spectrum radio galaxies. 

We set a lower limit for the luminosity of all the radio galaxies at $10^{23.5}$\,W\,Hz$^{-1}$. This limitation sets the number of flat-spectrum radio galaxies to 625.   We divided the steep-spectrum radio galaxies by NVSS luminosity into FR I and FR II samples. FR I galaxies have luminosity less than $10^{25}$\,W\,Hz$^{-1}$, and there are 793 of them in our sample. The luminosities of FR II galaxies are higher than this threshold and our sample contains 47 of them.

Seyfert galaxies were selected by the criteria presented by \citet{Kauffmann2003}. We required Seyfert galaxies to be detected photometrically as galaxies and to have emission lines [\ion{O}{iii}] $\lambda5007$, [\ion{N}{ii}] $\lambda6583$, H$\alpha$, and H$\beta$ with equivalent widths more than 3\,\AA. Galaxies with line ratios [\ion{O}{iii}]$/$H$\beta>3$ and [\ion{N}{iii}]$/$H$\alpha>0.6$ were selected as Seyfert galaxies. According to \citet{Hao2005}, the distribution of H$\alpha$ linewidths has a minimum at 1200\,km\,s$^{-1}$. We adopted this value as the limit between broad line Seyfert 1 and narrow line Seyfert 2 types. With these criteria, we get 1095 Seyfert 1 and 2494 Seyfert 2 galaxies.

\section{Methods \label{Methods}}

We study the environments of AGN in the supercluster scale by using a 
luminosity-density field constructed of LRGs. 
To see the supercluster-void network, the field is smoothed to 
appropriate scales. The smoothing length determines the characteristic scale of the structures under study \citep{Einasto2003}. A density field with a short smoothing length is suitable for studying cluster scales, while a longer smoothing length is needed for supercluster scales. We use a long smoothing length, so that voids, filaments, and superclusters have their own 
characteristic density levels, which can be used to distinguish between 
the details of the network.

To construct a luminosity density field, we assume that every galaxy is a visible member of a density enhancement, such as a group or a cluster. To compensate for the distance dependency of a magnitude-limited sample, we have to consider the luminosities of galaxies that drop out of the surveys magnitude window. Because of that the luminosities of galaxies are corrected by a weighting factor. This was done in the same way as in \citet{Lietzen2009}. 

The amount of unobserved luminosity is corrected by multiplying the observed luminosities of galaxies by weight factor $W_L(d)$, which is defined as
\begin{equation}
W_L(d)=\frac{\int_0^\infty \! L\phi(L) \, \mathrm{d} L}{\int_{L_1(d)}^{L_2(d)} \! L\phi(L) \, \mathrm{d} L},
\end{equation} 
where $L_1$ and $L_2$ are the luminosity limits of the galaxy sample and $\phi(L)$ is the galaxy luminosity function. The luminosity function was approximated by a double power law: 
\begin{equation}
n(L)\mathrm{d}(L)\propto(L/L^*)^\gamma)^{(\delta-\alpha)/\gamma}\mathrm{d}(L/L^*),
\label{Luminosityfunction}
\end{equation}
where $\alpha=-1.42$ is the exponent at low luminosities $(L/L^*)\ll1$, $\delta=-8.27$ is the exponent at high luminosities $(L/L^*)\gg1$, $\gamma=1.92$ is a parameter that determines the speed of the transition between the two power laws, and $L^*=-21.97$ is the characteristic luminosity of the transition \citep{Tempel2011}.

At distances over 400\,$h^{-1}$Mpc, the LRG sample is approximately volume-limited (number density is approximately constant) \citep{Liivamägi2010}. Because of this, weight correction is not needed in this region. At distances below 400\,$h^{-1}$Mpc (redshift 0.15), LRGs are fainter than LRGs at higher distances, and thus they are not real LRGs as defined by \citet{Eisenstein2001}, and they do not form a volume-limited sample \citep{Einasto2011}. However, they are similar to LRGs by many properties, and it is likely that they are distributed in the large-scale structure in the same way as the actual LRGs. We need the low-distance part of the LRG sample for comparing the LRG superclusters and the superclusters based on the main sample of galaxies. 

For the nearby LRGs, it is difficult to calculate the luminosity weights because the sample has no formal magnitude limits. In this region we found the observed comoving luminosity density $l(d)$ and defined the weight as
\begin{equation}
W_L(d)=l(d_0)/l(d),
\end{equation}
where $d_0= 435.6$\,$h^{-1}$Mpc ($z=0.15$) is used as a reference point, above which the weights are known from the luminosity function in Eq. \ref{Luminosityfunction}.  After correcting the luminosities we define a Cartesian grid for the whole survey volume, and calculate the luminosity density field on the grid using the B3-spline  kernel function. Details on constructing the density field are presented in \citet{Liivamägi2010}.

\begin{figure*}[ht]
  \resizebox{\hsize}{!}{\includegraphics{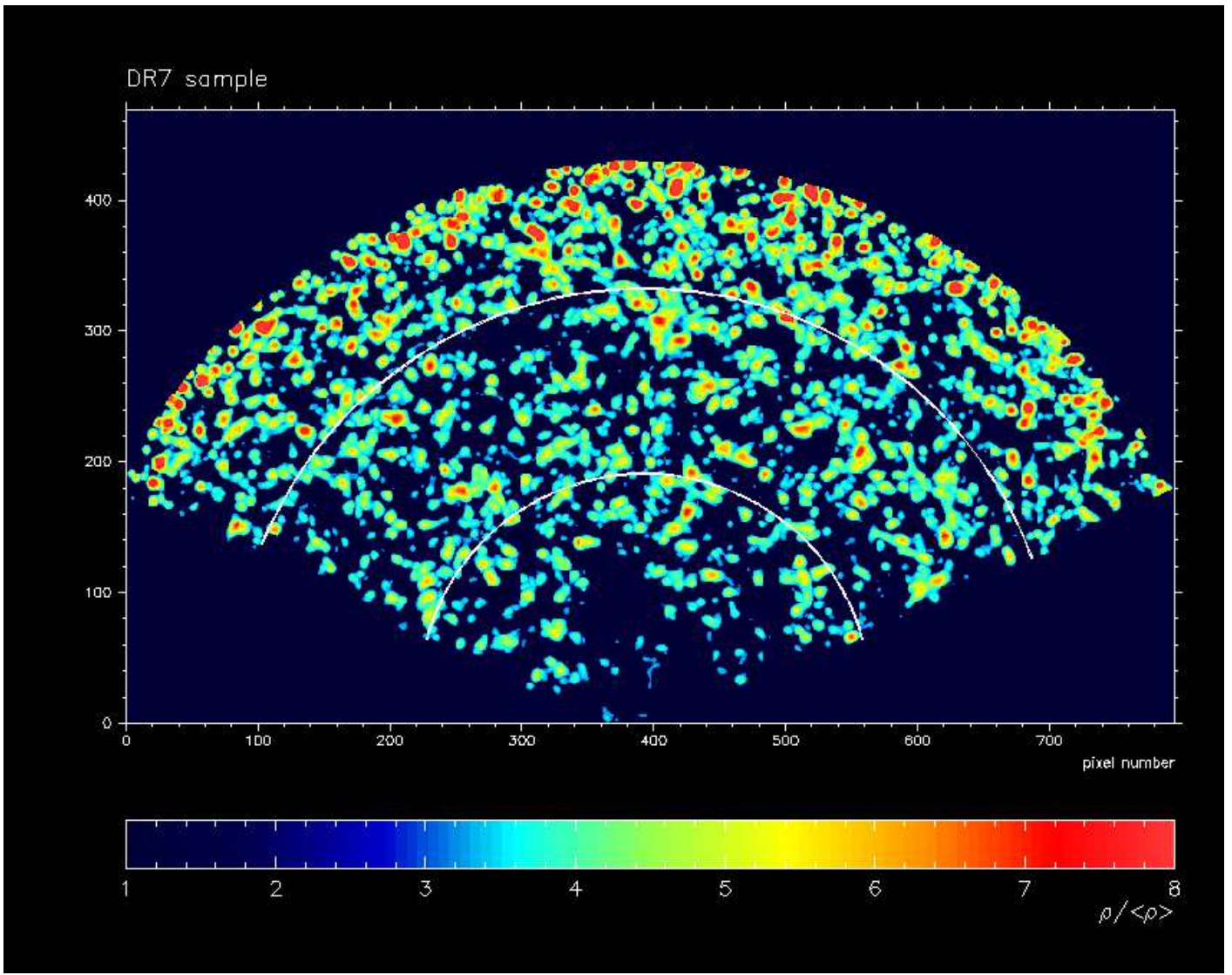}}
  \caption{Luminosity-density field of LRGs. The red spots are the brightest 
centers of superclusters, while the dark blue areas represent voids. The coordinates show the pixel number in the grid, 1 pixel corresponding to 3\,$h^{-1}$Mpc. The upper limits of the main sample of galaxies (565\,$h^{-1}$Mpc) and the part of the LRG field we use in our analysis (1000\,$h^{-1}$Mpc) are marked with white lines.}
  \label{LRGfield}
\end{figure*}

In this work we use a 3\,$h^{-1}$Mpc grid and 16\,$h^{-1}$Mpc effective radius of the smoothing kernel for the density field in our AGN environment study. 
Figure \ref{LRGfield} shows the LRG density field. In \citet{Lietzen2009} the smoothing kernel radius was 8\,$h^{-1}$Mpc. When using the LRG sample, a larger smoothing kernel is needed because the distances between the LRGs are longer than those between the main sample of galaxies. The volume of the LRG sample is considerably larger than the volume of the main sample of galaxies. The large size of the density field causes challenges to computing power and disk space. Because of this, we use a 3\,$h^{-1}$Mpc grid instead of the 1\,$h^{-1}$Mpc grid that was used in \citet{Lietzen2009}.

Although our LRG sample extends to a 1346\,$h^{-1}$Mpc distance, the  
weighting factors are too high for the highest distances, which cause densities that are too high at the farthest edge of the field \citep{Liivamägi2010}. These problems occur at distances greater than 1000\,$h^{-1}$Mpc. To remove the possible errors caused by the incorrect weighting at far distances, we set the 1000\,$h^{-1}$Mpc (redshift 0.4) as an upper limit for our analysis. We also set a lower limit for distance at 225\,$h^{-1}$Mpc because the volume, hence the number of AGN inclosed by this distance, is small.
Therefore, our sample extends from 225 to 1000\,$h^{-1}$Mpc. For a comparison to the LRG density field, we also construct a luminosity-density field of the main sample of galaxies using the same grid and smoothing kernel.
The field of main sample of galaxies we use extends from 225 to 565\,$h^{-1}$Mpc.

The lower limit for superclusters in the luminosity density field of the 
main galaxy sample with 3\,$h^{-1}$Mpc grid and 8\,$h^{-1}$Mpc smoothing kernel 
is 5.0 times the mean density \citep{Tempel2011}. This limit changes as we grow the kernel 
size to 16\,$h^{-1}$Mpc. We determine the new limit so that 
the summed luminosity of all grid points in the superclusters is the same 
as in the field with 8\,$h^{-1}$Mpc kernel. We get a limit of 
2.5 times the mean density for the field of the main sample of galaxies with 16\,$h^{-1}$Mpc kernel. The LRG sample consists of galaxies that are different from the main sample of galaxies.
Because of this, the luminosities of the main galaxy 
sample and the LRG sample are not unambigiously comparable, and we cannot use the luminosities as a comparison between these two fields. Instead, we determine the supercluster limit so that the number of density points in superclusters is the same in the LRG field as in the main galaxy field with 16\,$h^{-1}$Mpc kernel. In practice this means that the superclusters in the three-dimensional density fields of the LRG sample and the main sample overlap spatially. The supercluster limit we find for the LRG field this way is three times the mean density.

The limiting density for void areas in \citet{Lietzen2009} was 1.5 times the mean density. Following the same procedure as for the supercluster limit, we find that the limit corresponds to the density level of the mean density in the density fields with the 16\,$h^{-1}$Mpc smoothing kernel. The void limit is the same for both, the LRG and the main sample density fields. We define the regions with density values between the void and supercluster limit filaments. The same terminology was also used by \citet{Tempel2011}. The limiting densities for void, filament, and supercluster areas in different fields are shown in Table~\ref{DensityLevels}.
\begin{table*}
\centering
\begin{tabular}{c c c c}
\hline\hline
Field & Void & Filament & Supercluster \\
\hline
Main sample, 8\,$h^{-1}$Mpc kernel &$D<1.5$&$1.5<D<5.0$&$D>5.0$\\
Main sample, 16\,$h^{-1}$Mpc kernel&$D<1.0$&$1.0<D<2.5$&$D>2.5$\\
LRG,  16\,$h^{-1}$Mpc kernel&$D<1.0$&$1.0<D<3.0$&$D>3.0$\\
\hline
\end{tabular}
\caption{Definitions of voids, filaments, and superclusters in the luminosity-density fields (in units of mean density).}
\label{DensityLevels}
\end{table*}

For studying the environments of the AGN, we transform the coordinates of each AGN sample in the same 
cartesian coordinates as the density field. We then calculate the average 
density of the grid points of the density field at less than 3\,$h^{-1}$Mpc from each object (i.e. the nearest grid points surrounding the target). This 
gives us the local value of the low-resolution density field.

\section{Results}

We calculated the average density of the environment for our samples of 
quasars, BL Lac objects, radio galaxies, and Seyfert 1 and 2 galaxies. The quasars were divided into radio-quiet and radio-loud subsamples, and radio galaxies into flat-spectrum radio galaxies, steep-spectrum FR I galaxies, and steep-spectrum FR II galaxies as discussed in Sect. \ref{Data}. For comparison, we also calculated the densities in the environments of LRGs. We limited the LRGs between the distances 400\,$h^{-1}$Mpc and 1000\,$h^{-1}$Mpc. Between these distances the LRG sample is approximately volume-limited.  Table \ref{Results} shows the average density of the LRG density field for all objects of each type. It also shows the fraction of objects in void areas (less than 1 times the mean density), filament areas
(between 1 and 3 mean densities), and superclusters (more than 3 times 
the mean density). The average densities are shown in units of mean density of the whole density field, and the errors are standard errors of the average. 
\begin{table*}
\centering
\begin{tabular}{c c c c c c}
\hline\hline
Type & N & Average density & Void (\%) & Filament (\%) & Supercluster (\%) \\
\hline
Radio-quiet quasars& 3338&$1.71\pm 0.03$ &38&46&15\\
Seyfert 1 galaxies & 1095&$1.73\pm 0.04$ &34&51&15\\
Seyfert 2 galaxies&2494&$1.65\pm0.03$&35&52&13\\
Radio-loud quasars&26&$1.8\pm0.3$&35&42&23\\
BL Lac objects&81&$2.5\pm0.2$&26&42&32\\
Flat-spectrum radio galaxies&624&$2.60\pm0.07$&13&55&32\\
FR I radio galaxies&793&$3.01\pm0.07$&10&48&42\\
FR II radio galaxies&43&$3.2\pm0.4$&19&28&53\\
LRGs (nonactive) &75712&$2.862\pm0.007$&8&54&38\\
\hline
\end{tabular}
\caption{Number of objects, average environmental density of the LRG field, and percentage of objects in voids, filaments, and superclusters of different types of AGN between the distances from 225\,$h^{-1}$Mpc to 1000\,$h^{-1}$Mpc. }
\label{Results}
\end{table*}

\begin{figure*}[ht]
\centering
\begin{tabular}{c c}
  \resizebox{\columnwidth}{!}{\includegraphics{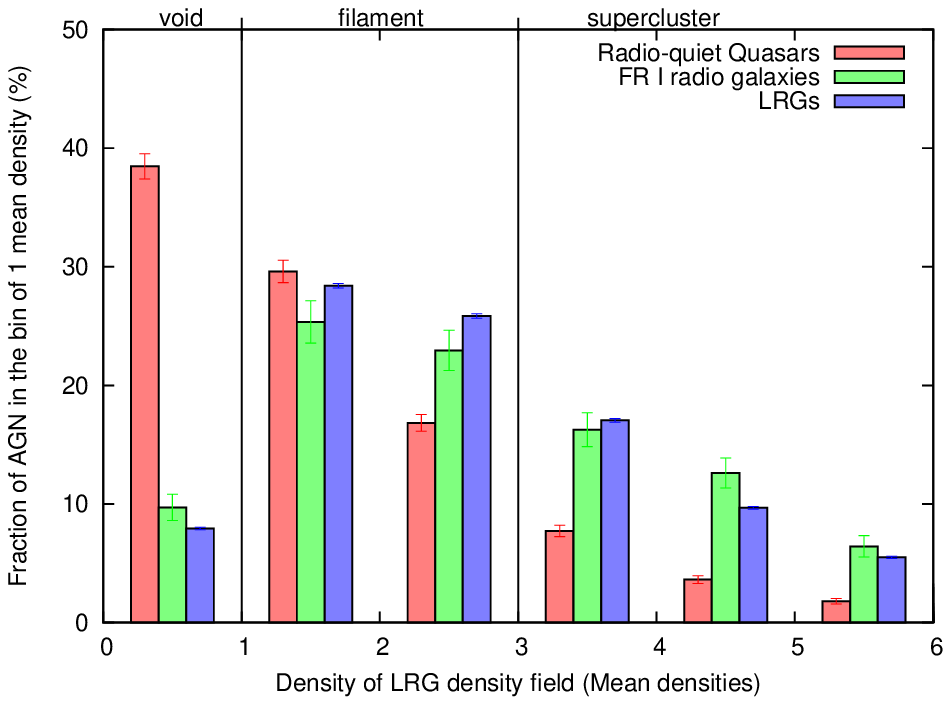}}&
  \resizebox{\columnwidth}{!}{\includegraphics{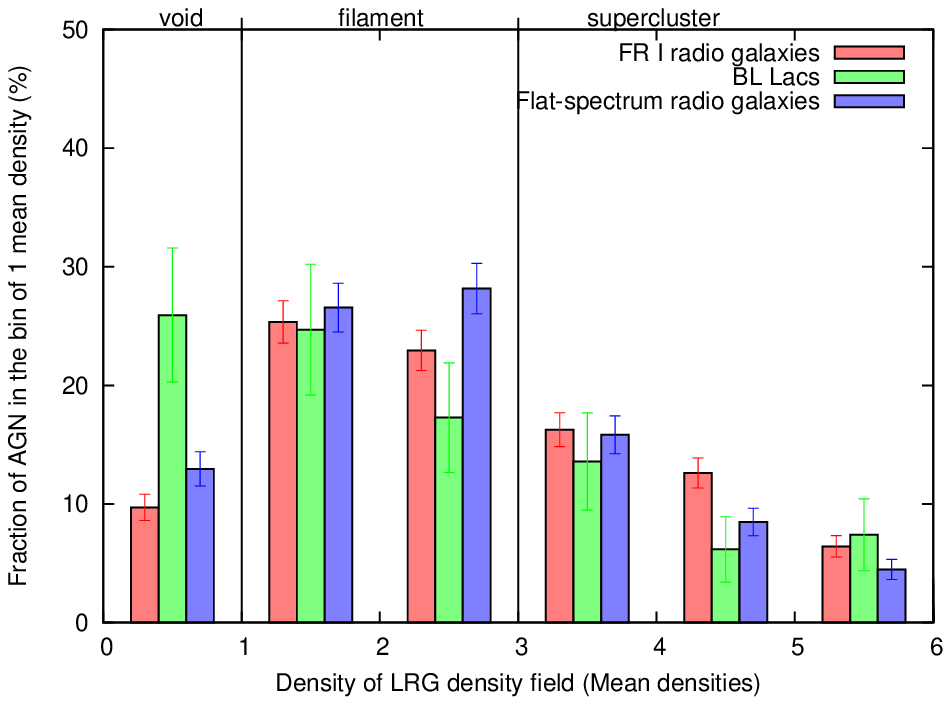}}\\
  \resizebox{\columnwidth}{!}{\includegraphics{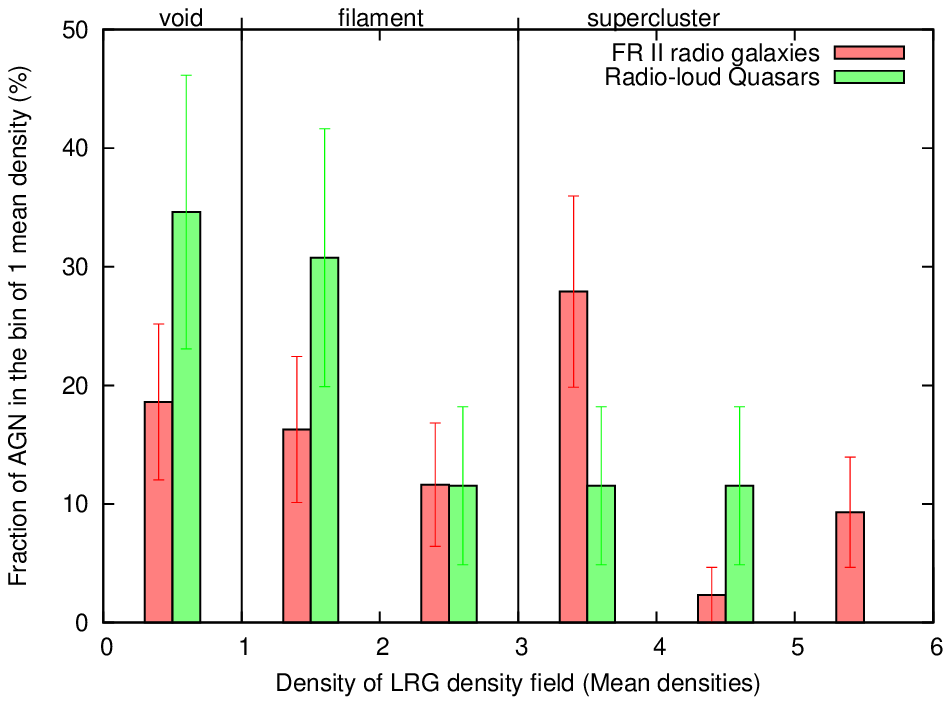}}&
  \resizebox{\columnwidth}{!}{\includegraphics{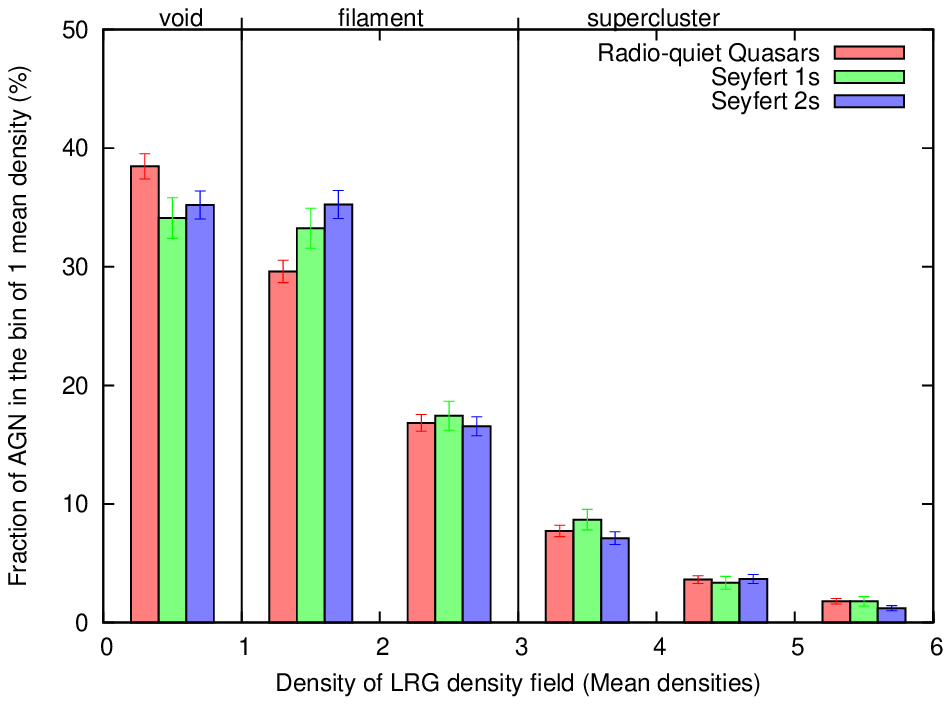}}
\end{tabular}
  \caption{Distributions of AGN in different levels of environmental density in the luminosity-density field of LRGs. The top-left panel shows the distributions of radio-quiet quasars, FR I radio galaxies, and LRGs. FR I radio galaxies, BL Lac objects, and flat-spectrum radio galaxies are shown in the top-right panel and FR II radio galaxies and radio-loud quasars in the bottom-left panel. The bottom-right panel shows the distributions of radio-quiet quasars and Seyfert galaxies.
Densities are shown in units of mean density of the density field. Error bars 
show Poissonian errors. The vertical lines separate the density levels of voids, filaments, and superclusters.}
  \label{DistributionsLRG}
\end{figure*}

Fractional distributions of the environmental densities of different types of AGN are shown in Fig. \ref{DistributionsLRG}.  The top-left panel gathers together  radio-quiet quasars, FR I radio galaxies and the non-active LRGs for comparison.  The top-right panel gives the distributions of FR I radio galaxies and BL Lac objects, which are thought to be similar objects viewed from different directions. The panel also plots the distribution of flat-spectrum radio galaxies. The bottom-left panel shows the distributions of radio-loud quasars and FR II radio galaxies, which make another unified group of AGN types. The bottom-right plot shows the distributions of radio-quiet AGN, i.e. Seyfert galaxies and radio-quiet quasars. Error bars in the histograms show Poissonian errors.

Table \ref{DistributionsGalaxy} shows the distributions of AGN in the 
luminosity-density field of the main sample of galaxies. This density field reaches the distance of 565\,$h^{-1}$Mpc. The average densities are shown in units of mean density of the whole density field, and the errors are standard errors of the average. The distributions in the LRG field with the same distance limit are shown in Table \ref{DistributionsNear}. The differences in the distribution of AGN in voids, filaments, and superclusters between the main sample field and the LRG field are small. This increases the credibility of using LRGs as indicators of the large-scale structures.
\begin{table*}
\centering
\begin{tabular}{c c c c c c}
\hline\hline
Type & N & Average density & Void (\%) & Filament (\%) & Supercluster (\%) \\
\hline
Radio-quiet quasars& 411&$1.52\pm 0.05$ &30&59&11\\
Seyfert 1 galaxies & 625&$1.58\pm 0.04$ &26&62&12\\
Seyfert 2 galaxies&1701&$1.51\pm0.02$&28&60&11\\
BL Lac objects&17&$1.9\pm0.3$&12&65&24\\
Flat-spectrum radio galaxies&189&$1.81\pm0.06$&11&72&17\\
FR I radio galaxies&174&$1.89\pm0.07$&14&66&20\\
\hline
\end{tabular}
\caption{Environmental properties of different types 
of AGN in the luminosity-density field of the main sample of galaxies at distances from 225\,$h^{-1}$Mpc to 565\,$h^{-1}$Mpc. }
\label{DistributionsGalaxy}
\end{table*}
\begin{table*}
\centering
\begin{tabular}{c c c c c c}
\hline\hline
Type & N & Average density & Void (\%) & Filament (\%) & Supercluster (\%) \\
\hline
Radio-quiet quasars& 411&$1.65\pm 0.07$ &36&52&12\\
Seyfert 1 galaxies & 625&$1.74\pm 0.05$ &33&53&14\\
Seyfert 2 galaxies&1701&$1.62\pm0.03$&35&53&12\\
BL Lac objects&17&$2.2\pm0.5$&24&53&24\\
Flat-spectrum radio galaxies&189&$2.19\pm0.09$&14&65&21\\
FR I radio galaxies&174&$2.3\pm0.2$&17&57&26\\
\hline
\end{tabular}
\caption{Environmental properties of different types 
of AGN in the luminosity-density field of LRGs at distances from 225\,$h^{-1}$Mpc to 565\,$h^{-1}$Mpc. }
\label{DistributionsNear}
\end{table*}

Our results for the environments in the LRG density field indicate that radio galaxies are more likely found in supercluster 
areas than radio-quiet quasars, which are mainly in low-density regions. 
Radio galaxies are actually distributed very similarly to LRGs themselves. This can be seen in the top lefthand panel of Fig. \ref{DistributionsLRG}. To test the similarities of different samples we performed the Kolmogorov-Smirnov test for our original unbinned results. Kolmogorov-Smirnov test confirms that the radio-quiet quasars and FR I radio galaxies are not from the same population at a level of more than 99.9\,\% significance.

The distribution of flat-spectrum radio galaxies is very similar to the distribution of steep-spectrum radio galaxies. This similarity can be seen in the top righthand panel of Fig. \ref{DistributionsLRG}, which shows the distributions of flat-spectrum radio galaxies and steep-spectrum FR I radio galaxies. This panel also plots the distribution of BL Lac objects, which are believed to be FR I radio galaxies viewed from the direction of the jet. This unification scheme predicts BL Lac objects to be in similar environments to FR I galaxies. The fraction of BL Lac objects in void regions is higher than for FR I galaxies, but the difference is small when considering the error limits. The Kolmogorov-Smirnov test gives a p-value of 0.06 for the likelihood that BL Lac objects and FR I galaxies are distributed similarly. This does not confirm that these samples are statistically different, but the higher fraction of BL Lac objects in the voids may suggest that at least some of them have different origins from the FR I galaxies.

The unification scheme of more luminous radio-loud AGN unifies radio-loud quasars with FR II radio galaxies. The distribution of these types in different environments is shown in the histograms plotted in the bottom lefthand panel of Fig. \ref{DistributionsLRG}. The radio-loud quasars in our sample show a slightly similar distribution to BL Lac objects: approximately 30\,\% of both of these types are in the voids, but also a considerable fraction is found in superclusters. Of the 26 radio-loud quasars, six are in supercluster areas. Their number is so small that random variation may play a role in their distribution. Because of this, we cannot say anything definitive about their distribution.  FR II galaxies are less usual in void areas, and seem to be distributed like FR I galaxies. This similarity may be caused by our making the division between FR I and FR II types based on only radio luminosity. We do not have any information on the morphology of our radio galaxies, which is the primary criterion of classification. The luminosity-based classification may be less certain in separating the physically different types. This may also happen when dividing quasars into radio-quiet and radio-loud samples. If radio-loud and radio-quiet AGN are of different origins, the division by luminosity is not necessarily as sharp as we have assumed. 

The difference between the distributions of FR II galaxies and radio-loud quasars looks large in Fig. \ref{DistributionsLRG}, but the small number of targets in these samples make the errors larger. The Kolmogorov-Smirnov test gives p-value of 0.03 for the similarity of radio-loud quasars and FR II galaxies.

The difference between the environments of radio-loud quasars and radio galaxies has also been detected by \citet{Donoso2010}. Using a cross-correlation function between AGN and LRGs with photometric redshifts, they find that radio-loud quasars are in lower density environments than radio-loud AGN. Our results support theirs even though we study AGN at lower redshifts, which limits the number of high radio-luminosity AGN in our sample, and thus causes uncertainty in our results. Our results on radio-loud and radio-quiet quasars are also quite similar to the results of \citet{Donoso2010}. There is no difference between the large-scale environments of radio-loud and radio-quiet quasars.

The distribution of Seyfert galaxies in different environments is close to that of radio-quiet quasars. We show their distribution in the bottom righthand panel of Fig. \ref{DistributionsLRG}. Despite their apparent similarity, the Kolmogorov-Smirnov tests between these samples give low p-values: the probability is 0.001 for the similarity of radio-quiet quasars and Seyfert 2 galaxies , and between Seyfert 1 and 2 types 0.02.

\begin{figure}[t]
  \begin{tabular}{c}
  \resizebox{\hsize}{!}{\includegraphics{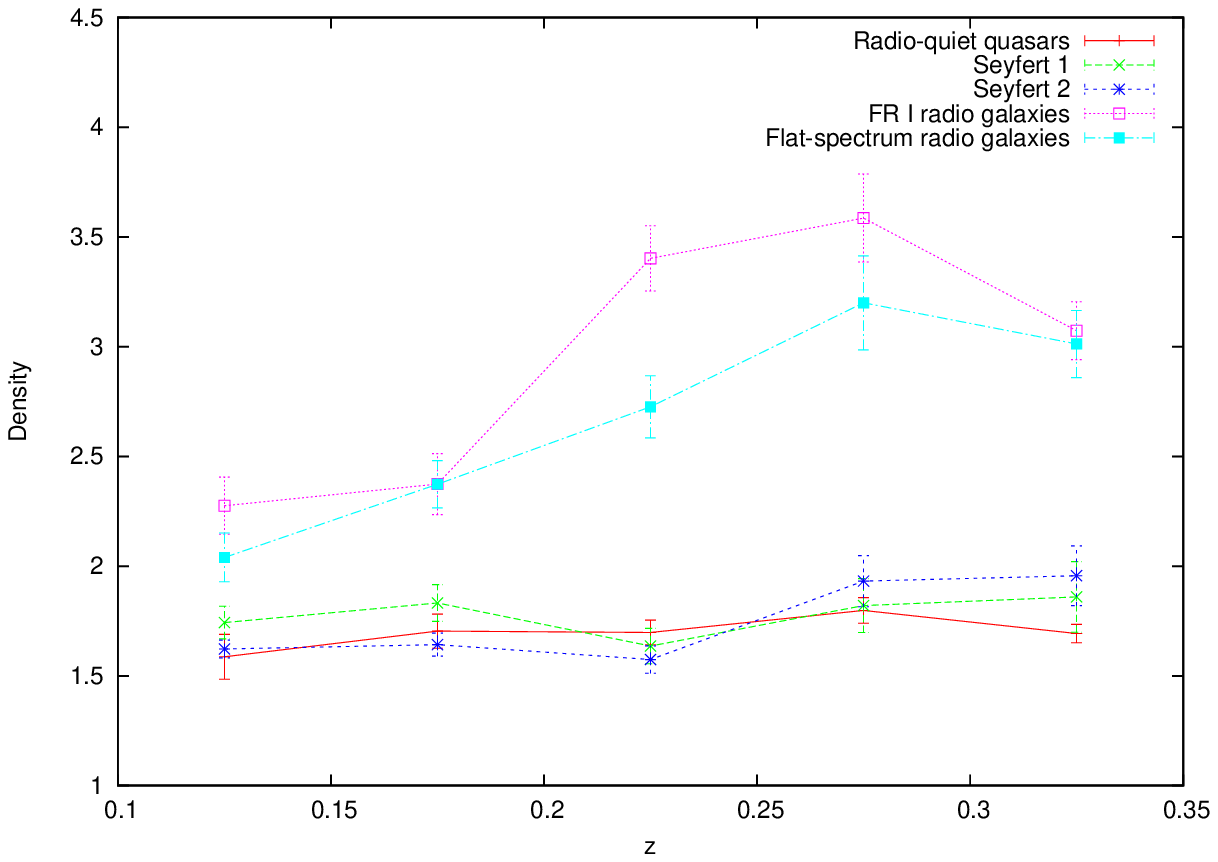}}\\
  \resizebox{\hsize}{!}{\includegraphics{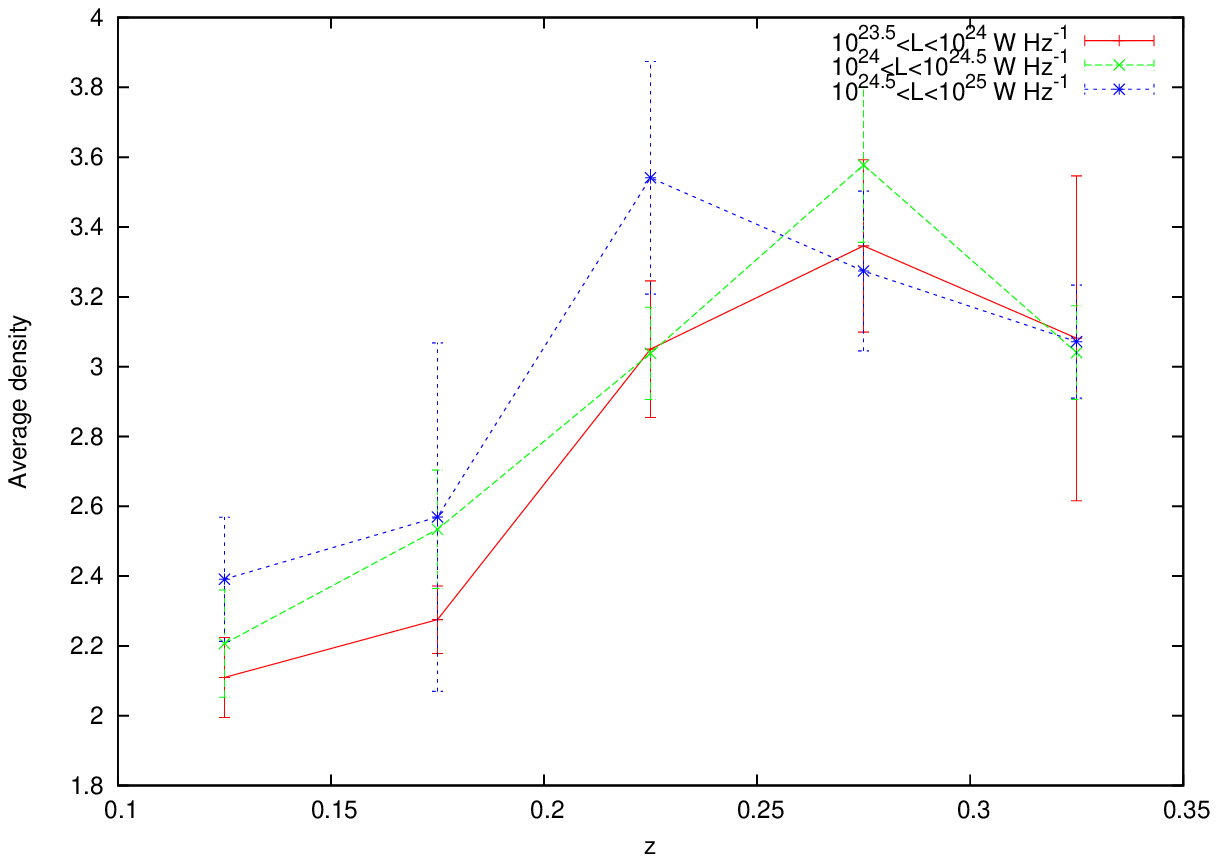}}
  \end{tabular}
  \caption{Top: Average environmental densities of radio-quiet quasars, Seyfert galaxies, FR I radio galaxies, and flat-spectrum radio galaxies as a function of redshift. Bottom: Average environmental densities of radio galaxies in three luminosity bins as a function of redshift. Densities are in units of mean densities and the errorbars show the standard errors of the average.}
  \label{zDependence}
\end{figure}

Figure \ref{zDependence} shows the redshift dependence of the average environmental density of different types of AGN. The top panel of the figure shows that the densities of radio-quiet AGN keep constant at different redshifts, while the radio galaxies at higher redshift are in higher density environments. This redshift-dependency could be caused by evolution but it could also come from selection effects. If high-luminosity radio galaxies are intrinsically in higher density environments than the fainter ones, a redshift dependency is observed because we only detect the most luminous galaxies at the highest redshifts. For quasars this is not a problem because the sample is volume-limited, and it should represent the true number density of quasars at all redshifts.  To test for possible observational selection effects we divided the radio galaxies into three luminosity bins: $10^{23.5}<L<10^{24}$\,W\,Hz$^{-1}$, $10^{24}<L<10^{24.5}$\,W\,Hz$^{-1}$, and $10^{24.5}<L<10^{25}$\,W\,Hz$^{-1}$. The bottom panel of Fig. \ref{zDependence} shows the dependence for radio galaxies in these three luminosity bins. Both steep and flat spectrum radio galaxies are included in this analysis to increase the data. Figure \ref{zDependence} shows that the relation between the density of the environment and redshift does not depend on the radio luminosity. We conclude that the radio luminosity does not cause a selection effect that would make us detect radio galaxies only in high-density environments at higher redshifts. Unlike radio galaxies, quasars have similar environments regardless of their redshift. We find no correlation between the absolute optical magnitudes or radio luminosities of quasars and their environments.

\section{Discussion}

Our results indicate that quasars are found in low-density areas in the 
large-scale structure of the Universe. This is consistent with the results in \citet{Lietzen2009}, as well as several other studies, such as
\citet{Padmanabhan2009} and \citet{Donoso2010}. 
Also our result that radio galaxies have denser enviroments than quasars is supported by \citet{Donoso2010}. Similarly, \citet{Hickox2009} found that radio-selected AGN are in higher density regions than X-ray or infrared-selected AGN. 

The high environmental densities of radio galaxies have also been found by \citet{Bornancini2010}. However, unlike \citet{Bornancini2010}, we do not find any difference between flat and steep-spectrum radio galaxies. We also did not find any luminosity dependence in the environments of radio-quiet AGN, like \citet{Strand2008} did. These differences may be due to the different scale.  Both \citet{Bornancini2010} and \citet{Strand2008} have studied the densities on the scale of an individual cluster of galaxies. Our method shows the average density on a large scale, and the situation may be different on the scale of individual clusters. 

On a larger scale, \citet{Mountrichas2009} determine an integrated correlation function up to scales of 20\,$h^{-1}$Mpc. They do not find any significant differences between quasars with different magnitudes. This supports our result that the large-scale environments of quasars do not depend on the optical magnitudes. The same method was used by \citet{daAngela2008} to study the magnitude redshift dependency of quasar environments. Also their results suggest that the environments of quasars do not change as a function of magnitude or redshift.

While the radio-quiet quasars are mostly located in low-density environments, radio galaxies are more likely in high density environments. The high density environments of radio galaxies was also detected on cluster scales by \citet{Wake2008}, who find that the central brightest cluster galaxies are often radio loud.

The unified scheme predicts that BL Lac objects would be in same kind of environments as FR I radio galaxies, while radio-loud quasars should have environments that are similar to FR II galaxies. Our results suggest that FR I and FR II galaxies are both likely to be in high-density environments. BL Lac objects are more often in low-density regions, but a larger fraction of BL Lacs than radio-quiet AGN are in superclusters. Since the number of  radio-loud quasars and FR II galaxies is very small, we cannot say anything conclusive about their unification. A more detailed study on the smaller scale environment of these objects could tell us more about whether they can be unified.

There is controversy about whether a difference exists between radio-loud and radio-quiet quasars. According to \citet{Shen2009}, radio-loud quasars have denser environments than radio-quiet quasars. On the other hand, \citet{Donoso2010} find no significant difference. Also our results show no dependency on the radio luminosity of quasars. \citet{Donoso2010} speculate that the disagreement with \citet{Shen2009} may have been due to different redshift ranges. When interpreting these results, one must bear in mind that  the error limits are large in all these studies. In any case, the difference is small, and detecting it may depend on the data and methods used.

We have made the first extensive study of the large-scale environments of 
BL Lac objects. We find that the BL Lacs in our sample were more evenly 
distributed at different environmental densities than other types of AGN. 
Although they are more common in low-density regions, a third of them are 
clearly in superclusters. The earlier studies of small-scale environments 
by \citet{Pesce2002} and \citet{Wurtz1997} support this, as their findings 
were also that some BL Lac objects are in rich clusters, while others 
are located in poor environments. \citet{Wurtz1997} also conclude that the environments of BL Lac objects resemble more quasars than FR I radio galaxies.
The density distribution we get on the supercluster scale seems to suggest that BL Lacs are not part of the same population as radio-quiet quasars, which are rarely in superclusters, but they do not avoid the void regions either, like the radio galaxies do. These results cause a conflict with the standard unification scheme that predicts BL Lac objects to be similar to FR I galaxies.

The differences between the environments of different types of AGN is 
supported by the properties of their host galaxies.
According to the morphology-density relation  
\citep{Dressler1997}, galaxies in dense environments are more often red 
ellipticals, while galaxies in less dense environments tend to be blue spiral 
galaxies. This was recently confirmed observationally by \citet{Lee2010}. They find that red galaxies have richer local environments than blue galaxies, and early-type galaxies have richer environments than late-type galaxies. Although the active nucleus may affect the general appearance of 
its host galaxy, our result seems to fit in this picture. According to 
\citet{Hyvonen2007}, BL Lac host galaxies are on average slightly  
redder than the host galaxies of quasars, but slightly bluer than radio galaxies. They also find that the colors of BL Lac host galaxies have a wider distribution than the other types of AGN. Radio galaxies are elliptical galaxies but have bluer colors than inactive elliptical galaxies \citep{Govoni2000}. Also \citet{Best2005} have found that radio galaxies are usually old, massive galaxies, while optically selected emission-line AGN are more spread out in host galaxy mass. According to \citet{Crenshaw2003}, Seyfert galaxies are nearly always spiral galaxies. These properties all fit  the morphology-density relation: radio galaxies are red ellipticals in the high-density environments, while Seyfert galaxies are spirals in the low-density environments. As \citet{Tempel2011} point out, the relation can be detected not only on cluster scale, but also in the global supercluster-scale environment. The location in the large-scale environment where a galaxy resides especially affects elliptical galaxies. Based on this background, our results for the environments of AGN give support to their role in galaxy evolution.

In Section \ref{Data} we raised questions about the nature of flat-spectrum radio galaxies. The flat-spectrum radio galaxies are distributed in different environments in the same way as the steep-spectrum FR I galaxies. This result speaks for the idea that they would be FR I radio galaxies whose radio luminosity is more concentrated near the core and then fades at farther distances. When they are observed at a high distance, the spectrum may look flat. 

We assume that the low-luminosity FR I radio galaxies may represent the `radio mode' feedback and quasars represent the `quasar mode' in the evolution model by \citet{Croton2006}. Based on this model, we would expect radio galaxies to originate in the steady inflow of gas from a static hot halo to the 
center of a massive galaxy, causing accretion into a central supermassive 
black hole. The efficiency of the accretion depends on the mass of the black 
hole and the virial velocity of the halo. This suggests that radio 
galaxies should be massive galaxies in massive clusters of galaxies, and 
therefore in high-density environments. As LRGs are massive galaxies, it 
is possible that they could be radio galaxies in a silent phase. This 
would explain why the environments of radio galaxies are similar to those of 
LRGs. Quasars, on the other hand, are expected to be found in environments 
where mergers between galaxies may have happened. According to 
\citet{Hopkins2008}, mergers occur in small groups of galaxies. This could 
explain why quasars have lower environmental densities than radio galaxies.

We found a redshift dependency in the environments of radio galaxies but not in the environments of quasars or Seyfert galaxies. This dependency does not depend on radio luminosity: radio galaxies of all luminosities have higher density environments at higher redshifts. This redshift dependence was detected on the cluster scale by \citet{Hill1991}, but they were unable to separate the possible effects of luminosity from those of redshift. We detect the same redshift dependency in the radio galaxies of all luminosities, which suggests that it is the redshift that causes the difference, not luminosity.

The results of \citet{Kauffmann2008} suggest that the majority of radio-loud AGN in the local universe are found in massive elliptical galaxies with weak or absent emission lines and little star formation. Through galaxy counts, they also find that radio galaxies reside in high-density environments on scales of a few hundred kpc and that the density does not depend on radio luminosity. They conclude that only low-mass black holes are growing and accreting at rates close to Eddington at the present day. At higher redshifts radio jets are stronger and radio galaxies are more likely to have powerful emission lines. \citet{Kauffmann2008} suggest that, since radio galaxies occur in halos with hot gas, while emission lines require cold gas in the surroundings of the galaxy, galaxies with both radio emission and emission lines need both. In the low-redshift universe, cold gas is available only in low-mass galaxies in low-density environments. This would explain the changes in radio-galaxy-population from high to low redshifts. Also \citet{Best2005} conclude that, based on their results on host galaxy properties, optically selected AGN and radio galaxies trace different populations of galaxies. Radio galaxies have more massive black holes and the galaxies themselves are more massive than optically selected AGN. Since massive galaxies are more commonly in high-density regions, this could explain why radio galaxies have high-density environments also on a large scale. 

In \citet{Lietzen2009} we speculated that low-redshift quasars have low-density environments because galaxies in dense environments evolve to a certain stage earlier in the Universe than galaxies in less dense regions \citep{Tempel2009}. Since quasars represent a relatively early phase in the life of a galaxy, the galaxies which show quasar-type activity at the present day have reached an early stage at a very late time. This would explain the low-density environments.  This is also suggested by \citet{Hickox2009}. If the radio galaxies are activated at a later phase in the evolution of galaxies than quasars, the low-redshift quasars we analyze have evolved slower than the radio galaxies at the same redshifts. Simulations by \citet{Gao2005} show that the dark matter clustering depends on the age of the halo. They conclude that the oldest 10\,\% of halos are more than five times more strongly clustered than the youngest 10\,\%. In the evolutionary model by \citet{Croton2006} the quasar mode turns the galaxies from blue star-forming galaxies into the red and dead ellipticals, while the radio mode regulates the growth of the large elliptical galaxies. This would place the radio mode a later phase in galaxy evolution than the quasar mode. Combining the predictions by \citet{Croton2006} and \citet{Gao2005}, we would expect that radio galaxies are old galaxies in dense environments. Quasars, on the other hand, are expected to be younger galaxies in less dense environments. 

\section{Conclusions}

In this paper, we used the luminosity-density field of LRGs for studying the global environments of different types of AGN. Our main result was that radio galaxies are more often in supercluster areas than radio-quiet quasars, which are more likely in void areas. This result can be explained by a model of galaxy evolution. Quasars are formed in mergers of gas-rich galaxies, and they lead to the formation of elliptical galaxies. In large elliptical, galaxies radio mode AGN activity regulates the growth. Radio galaxies are fed with hot gas, which is found in dense environments, while quasars are more likely to be in small dark matter halos that contain cold gas. In this scenario, the low-luminosity radio galaxies become active at a later phase in galaxy evolution than do quasars. As the superclusters are known to contain older galaxies than void regions, it is expected that the AGN activity more typical of older galaxies is more often found in superclusters than in voids. Therefore, our results give a portion of the much needed connection between the theoretical models of the effect of AGN activity in the evolution of galaxies and the actual observed AGN.

\begin{acknowledgements}
We thank the anonymous referee for comments that considerably helped to improve this article.

H. Lietzen was supported by the Finnish Cultural Foundation and the Academy of Finland project number 8122352. P. Hein\"am\"aki was supported by Turku University Foundation.

    Funding for the SDSS and SDSS-II has been provided by the Alfred P. Sloan Foundation, the Participating Institutions, the National Science Foundation, the U.S. Department of Energy, the National Aeronautics and Space Administration, the Japanese Monbukagakusho, the Max Planck Society, and the Higher Education Funding Council for England. The SDSS Web Site is http://www.sdss.org/.

    The SDSS is managed by the Astrophysical Research Consortium for the Participating Institutions. The Participating Institutions are the American Museum of Natural History, Astrophysical Institute Potsdam, University of Basel, University of Cambridge, Case Western Reserve University, University of Chicago, Drexel University, Fermilab, the Institute for Advanced Study, the Japan Participation Group, Johns Hopkins University, the Joint Institute for Nuclear Astrophysics, the Kavli Institute for Particle Astrophysics and Cosmology, the Korean Scientist Group, the Chinese Academy of Sciences (LAMOST), Los Alamos National Laboratory, the Max-Planck-Institute for Astronomy (MPIA), the Max-Planck-Institute for Astrophysics (MPA), New Mexico State University, Ohio State University, University of Pittsburgh, University of Portsmouth, Princeton University, the United States Naval Observatory, and the University of Washington.

\end{acknowledgements}

\bibliographystyle{aa}
\bibliography{references}
\end{document}